\documentclass[aps,prb,showpacs,floatfix]{revtex4}
\usepackage{graphicx}

\begin{document}

\title{Photoluminescence Spectroscopy  of   Band Gap Shrinkage  in  GaN}

\author{Niladri Sarkar and  Subhasis Ghosh} 

\affiliation{School of Physical Sciences, Jawaharlal Nehru University, New
Delhi 110067}

\begin{abstract}

We present an experimental investigation of band-gap shrinkage in n-type GaN  using photoluminescence
spectroscopy, as a function of electron concentration and temperature. The observed systematic
shift of the band-to-band transition energy to lower energies with increasing electron concentration   has been  interpreted as many-body effects due to exchange and correlation among
majority and minority carriers.   The band-to-band transition energy also shifts to lower energy  with increasing temperature. The parameters that describe the temperature dependence red-shift of the band-edge transition energy are evaluated using different models and we find that the semi-empirical relation based on phonon-dispersion related spectral function leads to excellent fit to the experimental data.

\end{abstract}

\pacs{78.20.-e, 78.55.-m, 78.55.Cr}
\maketitle

In recent years, GaN and its ternary alloys  have been extensively studied for their application in
blue and ultra-violet light emitting devices, short wavelength
lasers, and electronic devices for high power and high temperature
applications\cite{sn97}. This tremendous application potential
originates from the wide range of direct energy band gaps, the high breakdown
voltage, and the high saturation electron drift velocity.
In spite of technological breakthrough in GaN growth, doping and metal-GaN contacting technologies, several fundamental issues, such as temperature and doping induced band gap shrinkage(BGS) are  still controversial.   In particular BGS is  extremely important for the  development and modeling of GaN-based high temperature and high power devices. 
The effects of heavy doping on the optical and electrical
properties of semiconductors have been widely investigated\cite{raa78}
over the years because of two reasons: first the fundamental
physics underlying BGS and second their importance in
device applications. Heavily doped layers are integrated in
many devices and the estimation of BGS is an important
input in device simulations. 
Furthermore, several theoretical calculations\cite{gdm80,hsb81,bes86,cp02} with
different techniques have been published. Hence more experiments
on BGS in different semiconductors with different
intervals of doping would therefore appear motivated. The effect of  doping on the
band-to-band(BB) transition  energy in  heavily doped GaN has been studied by photoluminescence(PL)
and photoreflectance(PR) spectroscopy by several groups\cite{efs97,ms97,xz98,ei98,my99,ihl99} and the value of the BGS co-
efficient obtained from $n^{1/3}$ power-law fit varies between -$1.3\times10^{-8}$ to -$4.7\times10^{-8}eVcm$ at
room temperature, where $n$ is the electron concentration. Moreover, all the previous studies\cite{efs97,ms97,xz98,ei98,my99,ihl99} on GaN have ignored
the contribution of correlation term, which is proportional to  $n^{1/4}$ in
doping induced BGS. Recently, we have shown\cite{sg00} in detail the importance of this term on doping induced BGS in Al$_x$Ga$_{1-x}$As.

The  temperature induced BGS is observed  in experiments by a monotonic red shift of either BB and/or excitonic transitions  that are observed in bulk  as well as low dimensional  heterostructures. The  temperature dependence of the band gap $E_g(T)$ varies from relatively weak  in the low temperature  region to relatively strong at higher temperature region. There is considerable controversy regarding the temperature induced BGS and linewidth broadening.   All the previous studies\cite{ws95,mom96,kpk96,gdc96,cfl97,kbm04} on GaN have used empirical relations\cite{ypv67,lv84} neglecting the role of phonon dispersion in GaN\cite{jcn98,tr01}to explain  the experimental data. There is large variation  of the values of different BGS and linewidth broadening parameters, for example, in case of BGS, the most fundamental parameter   $\alpha$, which is  $-dE_g(T)/dT$  at high temperature limit,  varies  from    0.36meV/K to 1.2meV/K in the literature\cite{ws95,mom96,kpk96,gdc96,cfl97,kbm04}.  P\"{a}ssler\cite{rp99,rp02} has shown the inadequacy of these empirical relations for temperature induced BGS in semiconductors, in particular the inapplicability\cite{rp00} of these relations in case of wide band gap semiconductors. In addition,  residual  strain between GaN epilayers and substrates(sapphire and SiC) due to mismatch in lattice constant and thermal conductivity is also responsible for  scattering in the values of different BGS parameters.

PL spectroscopy has emerged as a
standard and powerful technique to study the optical properties
of bulk and low-dimensional semiconductor structures. It
is highly desirable to adopt a noncontact, nondestructive optical
technique to determine the homogeneity of the crystalline
quality and distribution of alloy composition in the epitaxial
layer. PL spectroscopy is routinely used to characterize
the quality of semiconductor substrates and thin epitaxial
layers for different device structures. 
PL spectroscopy is a direct way to measure the band-gap
energy as a function of different parameters, such as carrier
concentration, excitation intensity, temperature, pressure,
and magnetic field. The low-temperature PL spectra from
III-V binary and ternary compound semiconductors with reasonably
good crystalline quality are dominated by band-edge
and near-band-edge transitions. The underlying recombination
processes can be identified from the behavior of PL
spectra as a function of temperature and excitation intensity.
Low-temperature PL spectroscopy is usually performed to
study the excitonic recombination mechanisms in semiconductors,
but recently room-temperature PL spectroscopy has
received attention for different reasons. In most cases,
BB transitions can be observed only at room
temperature or at higher temperature. Several many-body effects
such as BGS, screening, and carrier-exciton scattering
can be studied using room-temperature PL spectroscopy.
Beside the understanding of basic physics regarding the
mechanism of recombination processes at higher temperature,
determination of the quantum efficiency of the BB transition
at room temperature is extremely important for optoelectronic
devices.  

In this work, we present the electron concentration and temperature dependence of BGS in GaN epilayers. In particular,  the role of background electron concentration on the band gap and different parameters responsible for temperature induced BGS   is studied. The results are compared with different empirical and semi-empirical models for BGS. 
Our samples are grown on c-plane sapphire substrates by
metal–organic chemical-vapor deposition(MOCVD) in a
vertical rf-heated, rotating disk quartz reactor at 76 Torr.
Previous studies\cite{yk00,ecp99} have shown that excess Ga in the buffer layer improves the crystal quality of GaN epilayers through stain relaxation and increased adatom surface mobility during the initial stage of epitaxial growth. We have recently shown\cite{sd02} that conductivity and the defect-related optical properties of GaN epilayers can be significantly controlled by the III/V ratio in GaN buffer layer. 
Prior to the growth of the epilayers, a 30-nm-thick GaN
buffer layer were grown at 565 °C. After growing the buffer
layer, the substrate temperature was raised to 1000 °C for
subsequent 1.6$\mu$m thick GaN epilayers. Five sets of
samples were grown with different trimethylgallium(TMG)
flow rates of 14, 24, 34, 48, and 55 mmol/min in the buffer
layer, while keeping the flow rate of NH$_3$ and H$_2$ constant at
1.5 and 2.5 slm, respectively.  The crystalline quality of the GaN epitaxial
films was evaluated by x-ray double-crystal diffraction
using GaN symmetrical(002) and asymmetrical(104)
reflections. More details of the samples are given in Ref.29. The electron concentrations were measured by
Hall measurements.   
  The electron concentrations were determined by Hall measurements.     We have chosen set of
 n-type samples with  electron concentration in the range of 
2.0$\times$10$^{16}$ to 2.7$\times$10$^{18}$ cm$^{-3}$.  The PL spectra
were collected in the wavelength region of 340-900nm. The samples
were kept in a closed cycle He refrigerator and were excited with
325nm laser line of  He-Cd laser. The PL signal was collected into
a monochromator and detected with a UV-enhanced Si detector.

Typical high temperature emission spectra in n-type GaN as a function of electron concentration are shown in Fig.1 and it is clear   that BB peaks shift towards lower energy as the electron concentration increases. This is a direct evidence for the BB transition. Fig.1.  also shows the PL spectra at 10K. The low temperature PL spectra show bound exciton(BE) at about 3.47eV  and multiple peaks due to donor-acceptor pair(DAP) recombination\cite{sd02,gdc96b,dv96} at lower energies.  We have avoided band-gap filling in our GaN samples by carefully choosing the carrier concentration such that except for the sample with carrier concentration   2.7$\times$10$^{18}$cm$^{-3}$, all samples are nondegenerate. It is clear in Fig.1  that BB peak is not broadened as the carrier concentration increases. The redshifts of BB transitions in GaN at 300K and 320K are shown in Fig.2.

In a heavily doped semiconductors, the density of states differs from that of the undoped pure crystal due to many-body effects. The presence of the large concentration of free carriers can cause a significant reduction of the unperturbed band gap in semiconductors. This reduction is caused by many-body effects and carrier-impurity interactions. As the doping density increases, the conduction band edge moves downward and the valence band edge moves upward. In addition to doping induced BGS, perturbation to the unperturbed band gap $E^{0}_{g}$ can come from (i) a random distribution of impurities which are almost independent of doping density\cite{kfb81}, and (ii) an interaction between electrons and donor impurities, which is appreciable only for very high carrier concentration($>10^{19}cm^{-3}$)\cite{kfb81}. It is almost impossible to separate the cumulative contribution to BGS from various many-body interactions. We are only interested in the role of electron-electron interaction on band gap. To avoid  contributions due to (i) and (ii), we have performed our BGS experiments with moderately doped samples with a maximum electron concentration of $10^{18}cm^{-3}$. The subject of this investigation
is to study the effect of electron-electron interactions
on BGS, which has been studied theoretically by several
groups\cite{gdm80,hsb81,bes86,kfb81} and 
  the effective change in the band gap due to many-body interactions can be given by

\begin{equation}
\Delta E_{g}=-an^{1/3}-bn^{1/4}
\label{eqn4}
\end{equation}

\noindent where, $a$ and $b$ are BGS co-efficients. The first term is due to the exchange interaction, which comes from the spatial exclusion of the like spins away from each other. The second term is due to the correlation energy, which comes from the repulsion of the like charges, so that they do not move independently, but in such a way so as to avoid each other as far as possible.  The values of co-efficients $a$ and $b$ are $1.5\times10^{-8}eVcm$ and $3.0\times10^{-9}eVcm^{3/4}$, respectively, evaluated by fitting with experimental data, as shown in Fig.2.

Temperature dependence of the band-gap $E_g$ of  two n-type GaN samples with different background electron concentrations is shown in Fig.3 and 4, which shows a maximum redshift of about 65meV as temperature increases from 10K to 320K.   
An appropriate fitting function is required to obtain material-specific parameters from experimentally measured  $E_g(T)$. There are two set of  fitting function for $E_g(T)$ in the literature: (i) empirical  relations proposed by Varshni\cite{ypv67} and Vi\~{n}a et al\cite{lv84}   and (ii) semi-empirical relations based  on  the electron-phonon spectral function $f(\epsilon)$  and the phonon occupation number $n(T)$,  $\epsilon$ is the phonon energy.

The most frequently  used empirical relation  for numerical fittings of E$_g(T)$ was first suggested by Varshni\cite{ypv67} and given by 

\begin{equation}
E_{g}(T)=E_{g}(0)-\frac{\alpha T^{2}}{\beta+T}
\label{eqn5}
\end{equation}

\noindent where E$_g(0)$ is the band gap at 0K, $\alpha$ is the T $\rightarrow \infty$ limiting value of the BGS coefficient $dE_{g}(T)/dT$ and  $\beta$ is a   material specific parameter.   This model represents a  combination of a linear high-temperature dependence with a quadratic low-temperature asymptote for $E_{g}(T)$. Though this phenomenological model  gave reasonable fittings of E$_g$(T)  in elemental,  III-V and II-VI semiconductors with E$_g$ $\leq$ 2.5eV, there have  been several problems with this relation, for example, (i) this relation gives negative values of $\alpha$ and $\beta$ in case of wide band gap semiconductors\cite{ypv67,rp00}, (ii)  $\beta$ is a physically undefinable parameter believed to be related to    Debye temperature $\Theta_{D}$ of the  semiconductor, but this connection has been doubted strongly in several cases\cite{rp02} and    (iii) it has been shown that this relation cannot  describe the experimental data of E$_g$(T) even in GaAs\cite{eg92}. 
 Fig.3(a) and 4(a) show the comparison of experimental E$_g$(T) with Varshni's relation\cite{ypv67}. We have obtained the values of $\alpha$ in the range of 0.54 meV/K to 0.63 meV/K and $\beta$ in the range of 700 K to 745K for  the  samples with  electron concentrations of 9.8$\times$10$^{16}$ to 2.7$\times$10$^{18}$ cm$^{-3}$. The parameter $\beta$ in this model only gives an estimation of  ${\Theta_{D}}$, which  is about 870K in GaN. It is clear that though the fitting is good in the low temperature region($<$ 100K), but fitting is poor in the intermediate($\sim$100K) and high temperature region($>$ 200K).

Vi\~{n}a et al\cite{lv84} first emphasized that total BGS, $\Delta E_g(T)=E_g(0)-E_g(T)$ is proportional to average phonon occupation numbers $\overline {n}(T)=\left[exp(\frac{\epsilon}{k_BT})-1\right]^{-1}$ and proposed an empirical relation, which can be expressed as 

\begin{equation}
E_{g}(T)=E_{g}(0)-\frac{\alpha_{B}\Theta_{B}}{\exp\left (\frac{\Theta_{B}}{T}\right)-1 }
\label{eqn6}
\end{equation}

\noindent where $\alpha_{B}=\frac{2a_{B}}{\bf{\Theta_{B}}}$, $a_B$ is a material parameters related to electron-phonon interaction,   $\Theta_B=\frac{\hbar\omega_{eff}}{k_B}$ represents  some  effective phonon temperature and $\omega_{eff}$ is the effective phonon frequency. Though this phenomenological model  gave reasonable fittings of E$_g(T)$  in different  semiconductors, there are also  several problems with this relation, for example, (i) at low temperature, this model  shows a plateau behavior of E$_{g}(T)$, which is not observed experimentally, (ii) at higher temperature($\geq$50K), this model predicts    $\Delta E_g(T)\propto exp(-\Theta_B/T)$, but, in most cases  $\Delta E_g(T) \propto T^2$ is observed experimentally,  (iii) it has been shown that this relation cannot describe the experimental data of E$_g(T)$ even in GaAs\cite{eg92}. Fig.3(b) and 4(b) show the comparison of experimental E$_g(T)$ with Vi\~{n}a's relation.  We have obtained the value of ${\Theta_{B}}$ in the range of 400K to 450K and  $\alpha_{B}$ in the range of 0.40meV/K to 0.44meV/K for  the  samples with  electron concentrations of 9.8$\times$10$^{16}$ to 2.7$\times$10$^{18}$ cm$^{-3}$. There is no doubt that this gives better fitting than Varshni's relation in the high temperature region, but the fitting in the low temperature region is poor.

As mentioned earlier, it can be  shown that the contribution of individual phonon mode to the temperature induced BGS is related to average phonon occupation number $\overline {n}(T)$ and electron-phonon spectral function $f(\epsilon$). Essentially,  $E_g$(T) can be analytically derived from the expression

\begin{equation}
E_{g}(T)=E_{g}(0)-\int d\epsilon {\it f}(\epsilon)\overline{n}(\epsilon,T)
\label{eqn7}
\end{equation}

\noindent The electron-phonon spectral function $f(\epsilon$) is not known {\sl a priori} and extremely complicated to calculate from the first principle. The other option is to use different approximate function for $f(\epsilon)$ to derive temperature dependence of BGS. 
It has been emphasized conclusively\cite{rp99,rp02} that the indispensable prerequisite for estimation of different parameters obtained from the experimentally measured $E_g(T)$ is the application of an analytical model that accounts for phonon energy dispersion. 
The BGS  results from the   superposition of contributions made by phonons with largely different energies, beginning from the zero-energy limit for acoustical phonons up to the cut-off energy for the optical phonons. The basic features of phonon dispersion $\delta_{ph}$ and the relative weight of their contributions to  $E_g(T)$, may vary significantly from one material to the other.
The curvature of the nonlinear part of the $E_{g}(T)$ is closely related to the actual position of the center of gravity, $\overline{\epsilon}$($\epsilon=\hbar\omega$), and the effective width  $\Delta\epsilon$, of the relevant spectrum of phonon modes that make substantial contribution to  $E_{g}(T)$ and this has been quantified by phonon-dispersion co-efficient $\delta_{ph}(=\frac{\Delta\epsilon}{\overline{\epsilon}})$.   It has been  shown\cite{rp00}  that  above two empirical models(Eqn.\ref{eqn5} and Eqn.\ref{eqn6}) represent the limiting regimes of either extremely large($\delta_{ph}\approx$1)   in case of Varshni's relation or extremely small($\delta_{ph}\approx$0) in case of Vi\~{n}a's relation. Both these models contradict physical reality  for most semiconductors, whose phonon dispersion co-efficient vary between 0.3 to 0.6\cite{rp00}. 
Several analytical models have been presented using  different forms of the spectral function ${\it f}(\epsilon)$.  P\"{a}ssler\cite{rp99} has proposed the most successful model, which takes  power-law type spectral function,  $f(\epsilon)=\nu\frac{\alpha_{p}}{k_{B}}\left(\frac{\epsilon}{\epsilon_{0}}\right)^{\nu}$
 and  the cut-off energy $\epsilon_{0}$, which  is given by $\epsilon_{0}=\frac{\nu+1}{\nu}k_{B}\Theta_{p}$.
Here, $\nu$ represents an empirical exponent whose magnitude can be estimated by  fitting the experimental data  $E_{g}(T)$.
Inserting, the spectral function ${\it f}(\epsilon)$ and cut-off energy $\epsilon_{0}$ into the general equation for band gap shrinkage(Eqn.\ref{eqn7}), P\"{a}ssler\cite{rp99} obtained an  analytical expression for $E_{g}(T)$, given by

\begin{equation}
E_{g}(T)=E_{g}(0)-\frac{\alpha_{p} {\bf{\Theta_{p}}}}{2}\left[\sqrt[p]{1+\left(\frac{2T}{\bf{\Theta_{p}}}\right)^{p}}-1\right]
\label{eqn8}
\end{equation}

\noindent where p=$\nu$+1 and  $\alpha_{p}$  is the T$\rightarrow \infty$ limit of the slope dE$_g$(T)/{dT}, ${\Theta_{p}}$ is comparable with the average phonon  temperature\cite{rp99},  ${\Theta_{p}}\approx \overline{\epsilon}/k_{B}$, and the  exponent p is related to the material-specific  phonon dispersion co-efficient $\delta_{ph}$, by the  relation $\delta_{ph}\approx1/\sqrt{p^{2}-1}$. Depending on the value of $\delta_{ph}$,  there are regimes of large and small dispersion which are approximately represented within this model by exponents $p<2$ and $p\geq3.3$ respectively. Fig.3(c) and 4(c) show the comparison of experimental $E_g(T)$ with Eqn.\ref{eqn8}.  We obtain an excellent fit to the experimental data.  
 The value of $\alpha_{p}$ obtained here is in the range of 0.50meV/K to 0.54meV/K and the value of ${\Theta_{p}}$  in the range of 500 K to 510 K for  the  samples with  electron concentrations of 9.8$\times$10$^{16}$ to 2.7$\times$10$^{18}$ cm$^{-3}$. The value of the  exponent $p$ which is related to the material specific degree of phonon dispersion is obtained between 2.50 to 2.65 for different GaN samples. The value of $p$ obtained here lies in the intermediate dispersion regime and the  value of the ratio  ${\Theta_{p}}/{\Theta_{D}}$ is 0.58, which means that the center of gravity of the relevant electron-phonon interaction are located within the upperhalf  of the relevant phonon spectra\cite{jcn98,tr01}. 
 
Fig.5 shows how  the total redshift $\Delta E_g=E_g(10K) - E_g(320K)$ and the phonon dispersion co-efficient $\delta_{ph}$  change  with electron concentration.     We have  observed that  $\Delta E_g$, which is the net change of E$_g$  due to the temperature induced BGS, decreases with increase of  electron concentration. This has been explained  as the screening of the  electron-phonon interaction which causes temperature induced BGS by background electrons. When the electron concentration is high, $\Delta E_g$ will be less as compared to the case when electron concentration is low.  Now the similar effect  will also be reflected in the E$_g$ vs. T curves. At lower electron concentration, the curvature  of the nonlinear part of the $E_g(T)$ will be more. Hence the value of the exponent p which is obtained from fitting our experimental data will vary with electron concentration. The value of p which controls the curvature of the nonlinear part of the $E_g(T)$ will be less at lower electron concentration and will increase at higher electron concentrations. Now, as mentioned earlier, p is inversely related to  $\delta_{ph}$, by the relation $\delta_{ph}\approx1/\sqrt{p^{2}-1}$. Hence the value of  $\delta_{ph}$ will decrease as electron concentration increases. The inset of Fig.5 shows this.

In conclusion, BGS  has been studied in GaN as a function of electron concentration
and temperature. The band-to-band transition
shows a redshift due to band-gap narrowing as the electron
concentration  increases. It has been
found that exchange and correlation among majority and minority
carriers causes band-gap narrowing in moderately
doped n-type GaN. We have also  measured the temperature induced band gap reduction in GaN using PL spectroscopy.   The importance of electron-phonon interaction on the band gap shrinkage has been established.   It has been found that phonon-dispersion based semi-empirical relation is required to explain the experimental data.  Screening of electron-phonon interaction is responsible for decrease of temperature induced BGS in samples with higher electron concentration.

The authors thank Professor J. H. Edgar of Kansas State
University, for providing some of the  samples used in this investigation, and for helpful discussions
and encouragement. This work was partly supported by Council of Scientific
and Industrial Research, India.

\newpage

\noindent {\large\bf Figure Captions}

\vspace{0.5in}

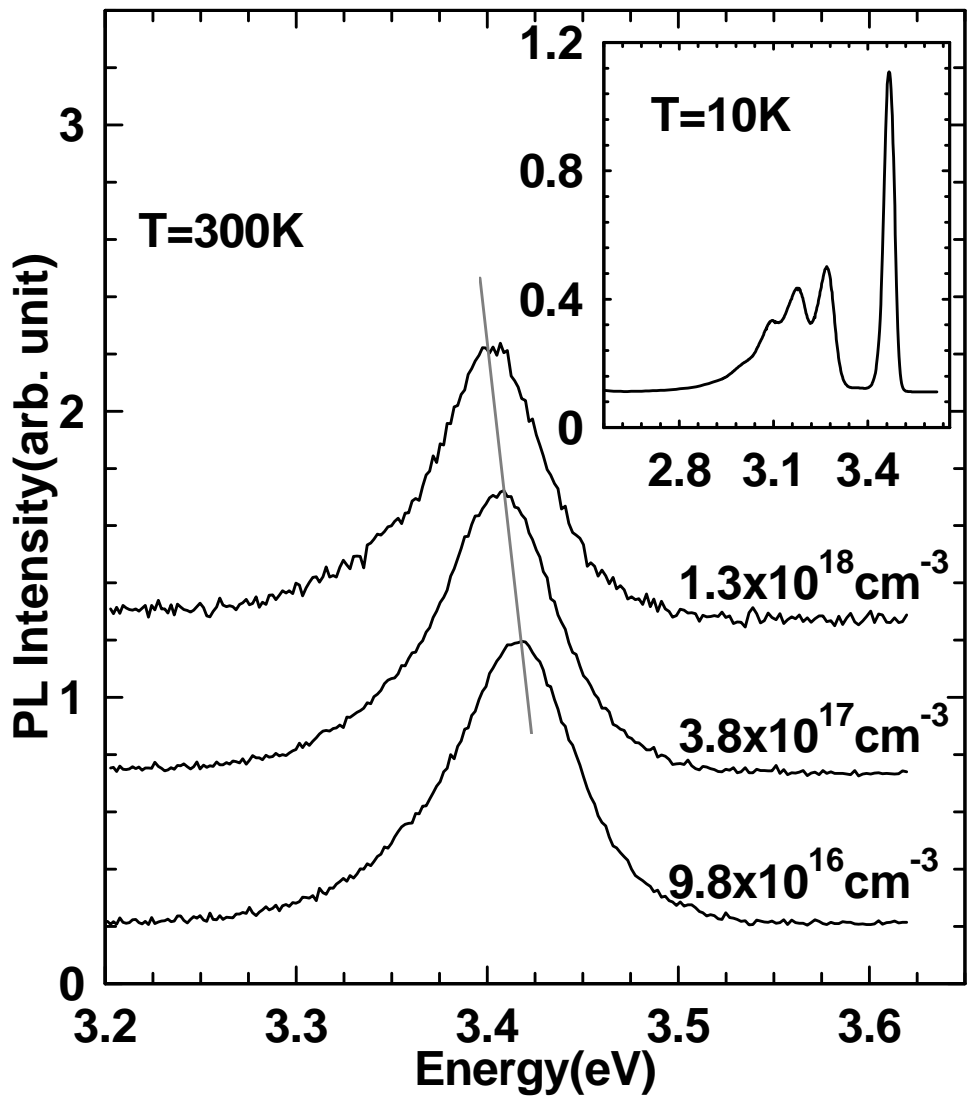
\noindent Figure 1.  PL spectra of GaN with different   electron concentrations
at 300 K. An increase of the doping level shifts the band-to-band
peak to lower energies.  PL  spectra are shifted up for clarity. PL spectrum of GaN with 3.8$\times$10$^{17}$cm$^{-3}$ electron concentration at 10 K is shown in the inset. 

\vspace{0.5in}

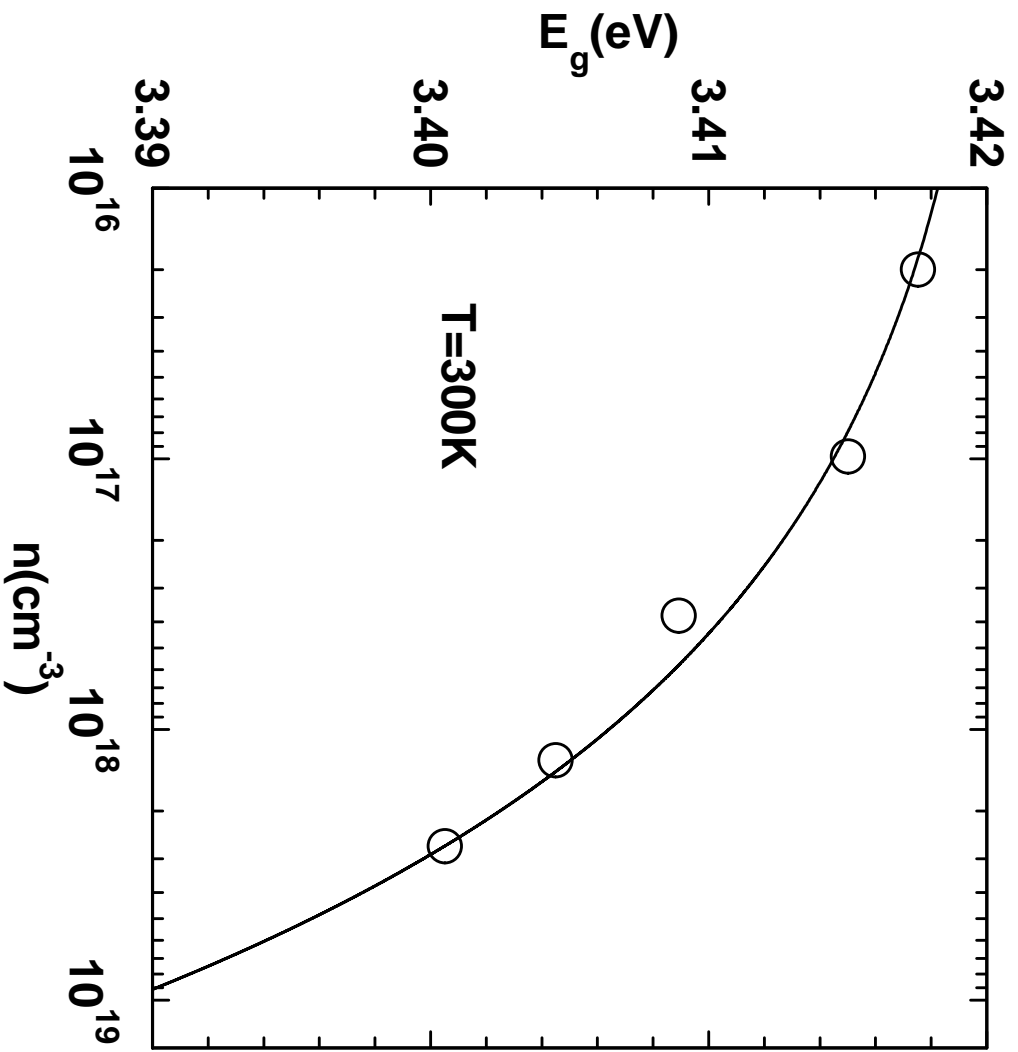
\noindent Figure 2. Band-gap energy determined from the  PL spectra
 shown as a function of logarithm of electron
concentration. Solid
lines are the fitted curves to the Eqn.1.

\vspace{0.5in}

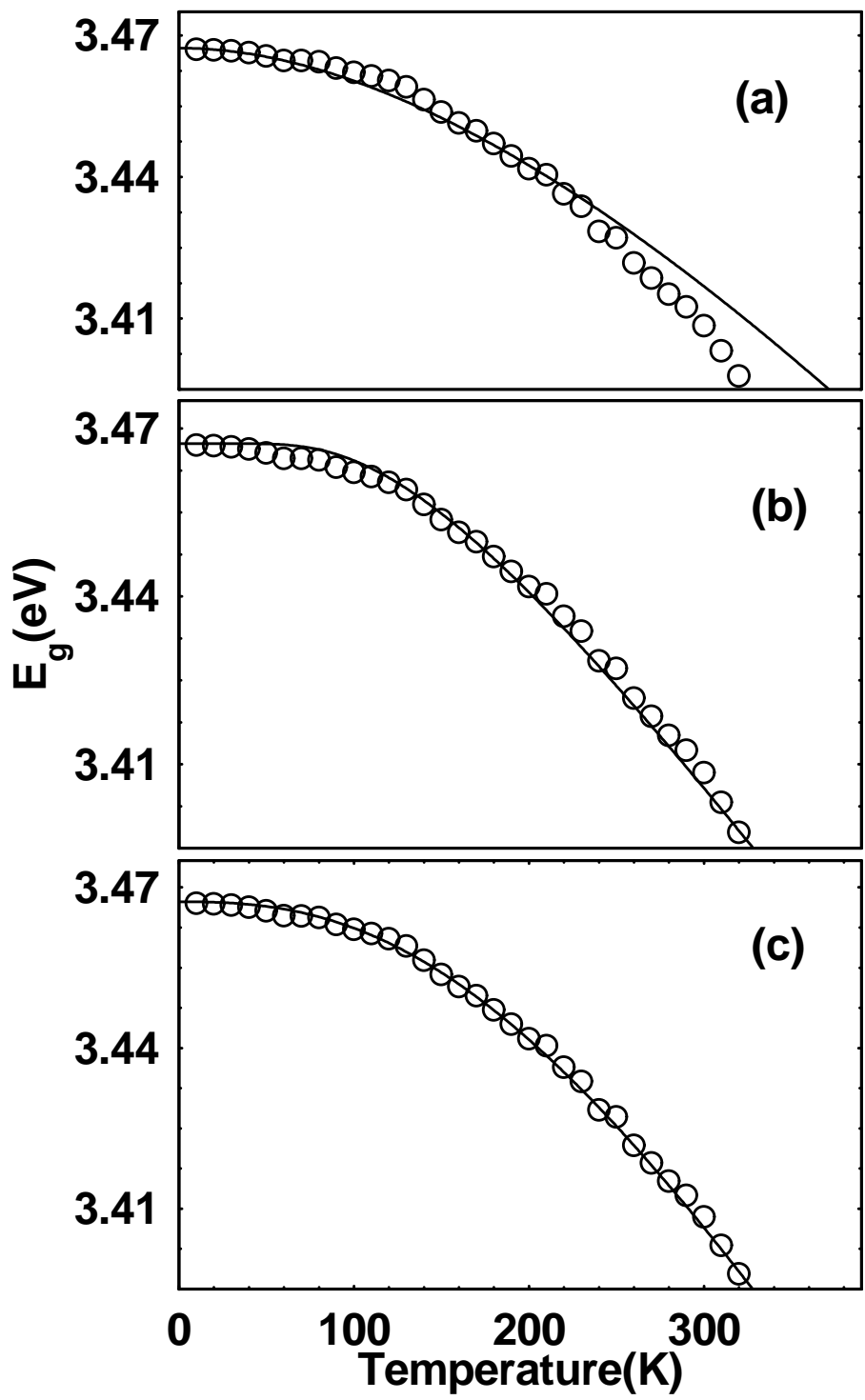
\noindent Figure 3. Temperature dependence of the peak positions of band-edge excitonic transition in GaN sample with electron concentration of  3.8$\times$10$^{17}$cm$^{-3}$.    Solid lines are fit to experimental data with (a) Eqn.2, (b) Eqn.3 and (c) Eqn.5.

\vspace{0.5in}

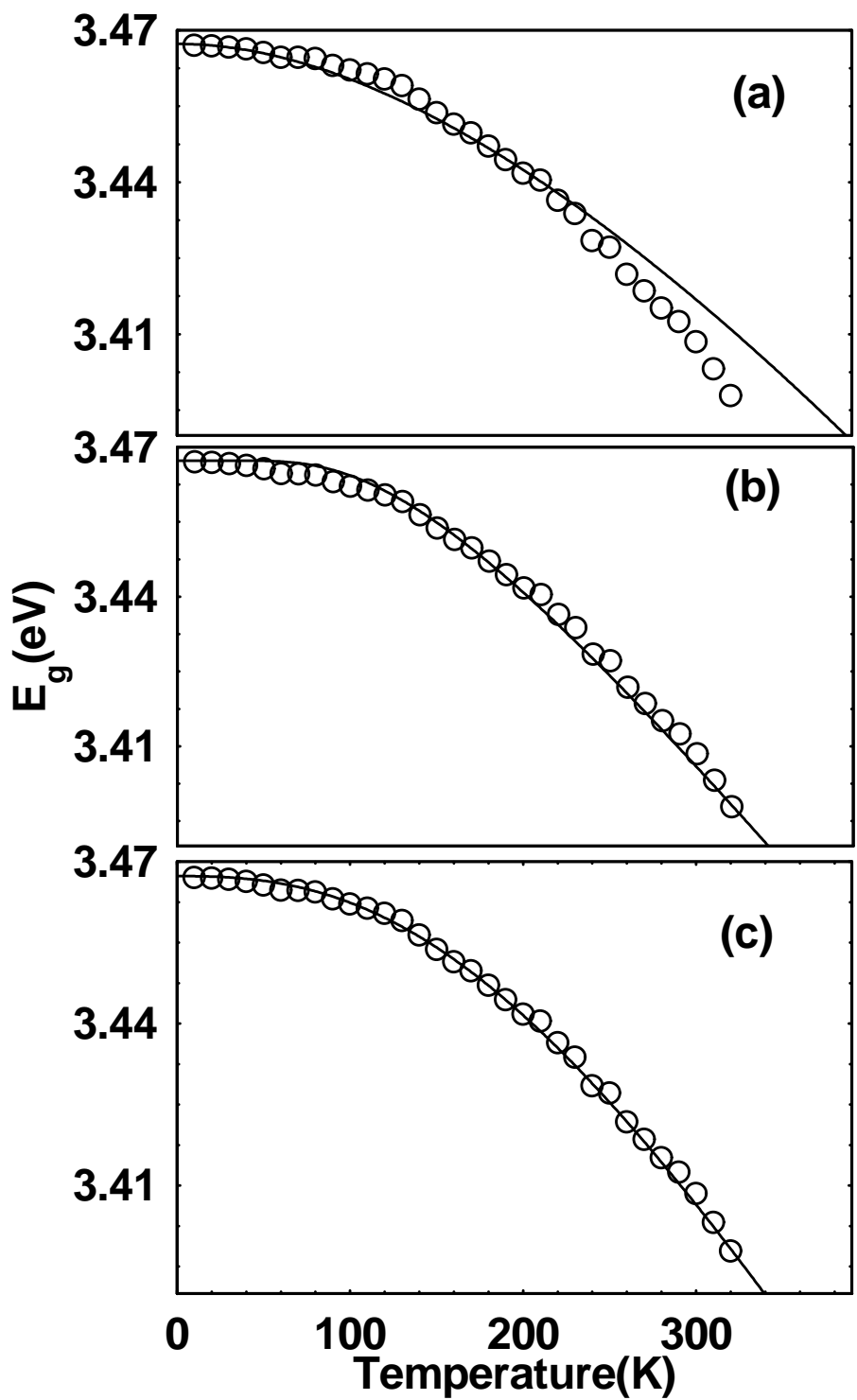
\noindent Figure 4. Temperature dependence of the peak positions of band-edge excitonic transition in GaN sample with electron concentration of  1.3$\times$10$^{18}$cm$^{-3}$.    Solid lines are fit to experimental data with (a) Eqn.2, (b) Eqn.3 and (c) Eqn.5.

\vspace{0.5in}

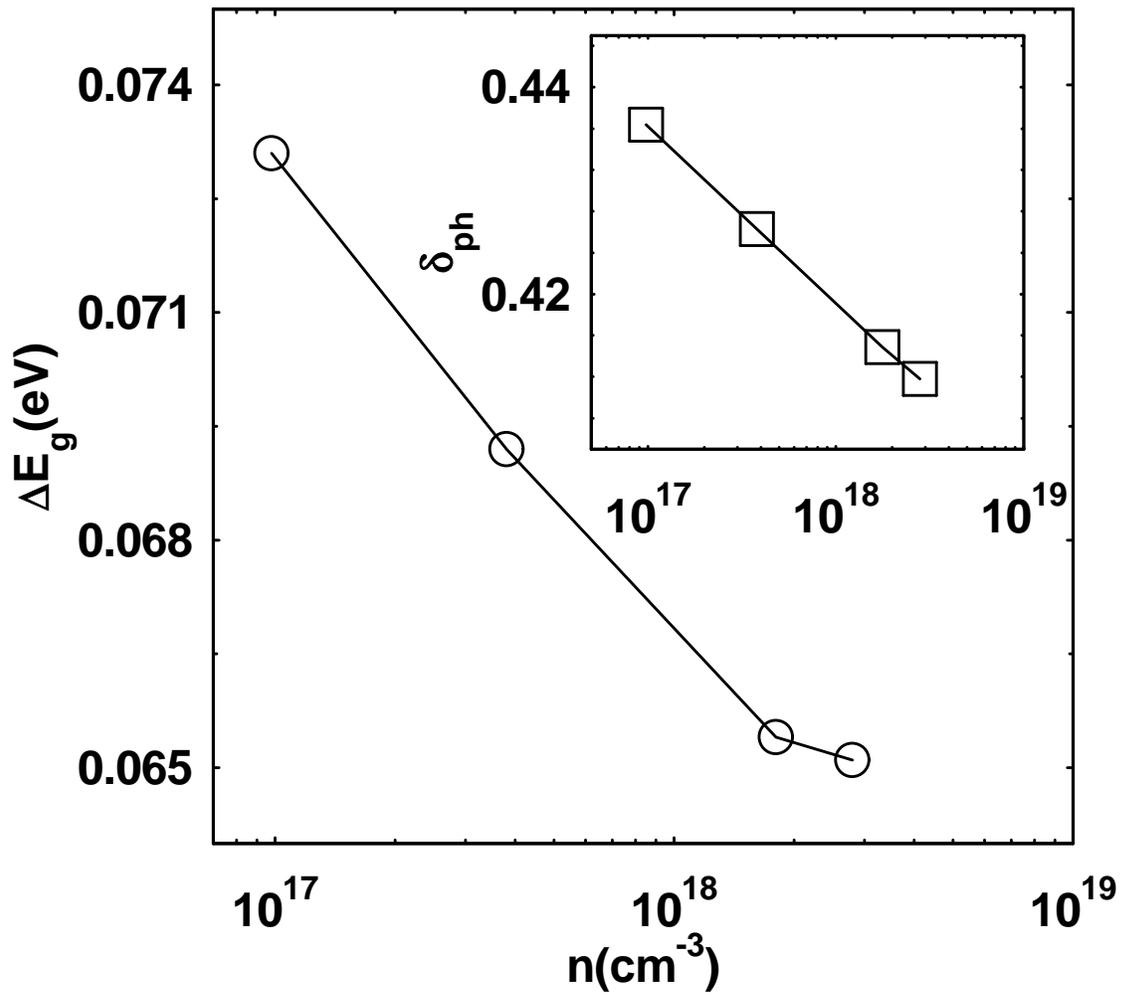
\noindent Figure 5. Electron concentration dependence of BGS, $\Delta E_g=E_g(10K) - E_g(320K)$. Inset shows how the phonon-dispersion parameter $\delta_{ph}$ derived from relation $\delta_{ph} = 1/\sqrt {p^2-1}$  changes with electron concentration $n$. Solid lines are guide for eyes.

\end{document}